\documentstyle[12pt]{article}
\begin{document}
\bibliographystyle{unsrt}
%%%%%%%%%%%%%%%%%%%%%%%%%%%%%%%%%%%%%%%%
\newcommand{\Po}{Poincar\'e}
%%%%%%%%%%%%%%%%%%%%%%%%%%%%%%%%%%%%%%%%%%%%%%%%%%%%%%%%%%%%%%%%%%%%%%%%%%%
\title{\huge\rm Conceivable new recycling of nuclear waste by nuclear power
companies \\ in their plants}
%%%%%%%%%%%%%%%%%%%%%%%%
\author{Ruggero Maria Santilli\\*[1ex]
{\normalsize\it Institute for Basic Research, P.O. Box 1577, Palm Harbor,
FL-34682, USA}}
\date{}
\maketitle
%%%%%%%%%%%%%%%%%%%%%%%%%%%%%%%%%%%%%%%%%%%%%%%%%%%%%%%%%%%%%%%%%%%%%%%%
%%%%%%%%% J&rgneb Los Alamose keskuse n6uete kohane preambula: %%%%%%%%%

%\\
%Title: Conceivable new recycling of nuclear waste by nuclear power
%companies in their plants
%Author: Ruggero Maria Santilli (Institute for Basic Research, P.O. Box 1577,
%Palm Harbor, Fl-34682, USA; e-mail:ibr@gte.net)
%Comments: ... pages, LaTex,
%........... (????)
%Report-no: ??
%Subj-class: PACS 03.65.-w;21.10.Hw;21.10.Ky
%Journal-ref: ??
%%%%%%%%%%%%%%%%%%%%%%%%%%%%%%%%%%%%%%%%%
%Abstract:
%\\
{\bf Abstract.} We outline the basic principles and the needed experiments
for a conceivable new recycling of nuclear waste by the power plants
themselves to avoid its transportation and storage to a (yet unknown) dumping
area. Details are provided in an adjoining paper and in patents pending.
%\\
%%%%%%%%%%%%%%%%%%%%%%%%%%%%%%%%%%%%%%%%%%%%%%%%%%%%%%%%%%%%%%%%%%%%%%%%%
%%%%%%%%%%%%%%%%%%%%%%%%%%%%%%%%%%%%%%%%%%%%%%%%%%%%%%%%%%%%%%%%%%%%%%%%%
\newpage

\section{Introduction}

The recycling of nuclear waste is certainly one of the largest open problems
of contemporary science, industry and society. Current studies (see, e.g.,
[1a]) are essentially conducted under the assumption of first transporting
the waste to a (yet unknown) dumping area, and then recycling it, e.g.,
via the construction of a huge particle accelerator for the individual
smashing of the waste nuclei (see, e.g., [1b]).

Even though technically possible and quantitatively predictable on grounds
of current knowledge, this approach is not immune from problematic aspects,
such as: the high cost of first transporting and storing the waste and then
recycling it (estimated in the range of hundred of {\sl billions} of dollars
in the U.S. alone [1a], with similar expenditures predicted for Europe);
the danger to the public caused by the fact that the transportation of the
highly radioactive waste can only occur via ordinary trucks traveling in
ordinary streets (thus being predictably opposed by consumer groups); the
long time needed for the above recycling (estimated in the range of hundred
of years due to the thousands of tonnages of radioactive waste to be
recycled); etc.

In this note we initiate the study of novel means of recycling nuclear waste
specifically conceived to be usable by the nuclear power companies in their
own plants, e.g., in their water pools, thus avoiding altogether the costly
and dangerous transportation of the waste to a dumping area. The new
recycling might also stimulate a new industry for the development and
production of the needed new equipment for all interested nuclear power
plants.

Due to the practical impossibility of building large particle accelerators
inside nuclear power plants, the only possible alternative is to identify
means capable of disrupting the nuclei of the waste via resonating or other
mechanisms, so as to drastically reduce meanlives from current values of the
order of thousands of years, down to values of the order of weeks or days.

It should be indicated from the outset that the latter means {\sl are not}
predicted by relativistic quantum mechanics (RQM) and its underlying
Poincar{\'e} symmetry $P(3.1)$ because, in their realization via the familiar
Dirac equation, the latter structures predict perennial and immutable
meanlives. Thus, quantitative studies of the desired new forms of recycling
nuclear waste ``in house'' require a suitable broadening of RQM and related
Poincar{\'e} symmetry.

In this note we recall that RQM is {\sl exactly} valid for the atomic
structure but {\sl not} for the nuclear structure with predicted small
deviations of crucial relevance for the desired recycling; we point out a
broadening of RQM known under the name of {\it relativistic hadronic
mechanics} (RHM) with corresponding broadening of its basic {\Po} symmetry,
specifically constructed for the nuclear structure with extended and
deformable nucleons; we point out that the alterability of the {\it intrinsic}
magnetic moment of nucleons is a necessary condition for the representation
of total nuclear magnetic moments and we present its first exact-numerical
representation via RHM under conventional values of spin and angular
momentum; we indicate that the alterability of the intrinsic magnetic moments
of the neutron implies the alterability of its meanlife via {\it subnuclear}
processes; we present the basic principle of the proposed new recycling
of nuclear waste based on subnuclear resonating mechanism which stimulate the
beta decay of the neutron; and we finally point out three fundamental
experiments which are essential for additional studies in the proposed new
recycling.

A detailed presentation is available in the adjoining paper [1c] and in
patents pending. The reader should be warned against unreasonable
expectations of quick solutions in view of the magnitude and complexity of
the problem whose solution will predictably require a comprehensive
collegial effort encompassing all current and new scientific and industrial
knowledge in nuclear physics.

%%%%%%%%%%%%%%%%%%%%%%%%%%%%%%%%%%%%%%%%%%%%%%%%%%%%%%%%%%%%%%%%%%%%%%%%%%%%
\section{Lack of exact character of relativistic \hfil\break
quantum mechanics in nuclear physics}

RQM resulted to be exact for the atomic structure because it represented
in an {\sl exact way} all its experimental data. By comparison, RQM
{\sl cannot} be {\sl exact} for the nuclear structure because it has been
unable to provide an {\sl exact} representation of {\sl all} its experimental
data, such as total magnetic moments and other data [2]. The understanding is
that RQM does indeed provide a good {\it approximation} of nuclear data.
However, as we shall see, even though evidently small, the expected
deviations have a fundamental role for the desired new recycling of nuclear
waste.

The lack of exact character of RQM in nuclear physics can also be seen in a
number of independent ways, a compelling one being that based on symmetries.
Computer visualization of the {\Po} symmetry indicates its capability to
represent {\it Keplerian systems}, i.e., systems with the heaviest
constituent in the center, as occurring in the atomic structure. By
comparison, {\sl nuclei do not have nuclei} and, therefore, the {\Po}
symmetry must be broken to represent structures {\sl without} the Keplerian
center. In turn, as we shall see, the latter breaking is fully in line with
the deviations from $P(3.1)$ required by the representation of nuclear
magnetic moments.

The most compelling arguments are of {\it dynamical} nature. RQM was
constructed for the characterization of action-at-a-distance interactions
derivable from a potential, as occurring in the atomic structure. By
comparison, nucleons in a nuclear structure are in an average state of
mutual penetration of about $10^{-3}$ parts of their charge distribution
[3a]. But hadrons are some of the densest objects measured in laboratory
until now. Their mutual penetration therefore implies a (generally small)
component of the nuclear force which is: 1) of {\it contact} type i.e.,
with zero-range, thus requiring new interactions {\sl without} particle
exchanges; 2) {\it nonlinear} in the wave functions and, possibly, their
derivatives, thus requiring a theory with an exact superposition principle
under said nonlinearity; 3) {\it nonlocal}, e.g., of integral-type over the
volume of overlapping, thus requiring a new topology; 4) {\it nonpotential},
in the sense of violating the conditions to be derivable from a potential
or a Hamiltonian [3b], thus requiring new dynamical equations; and 5) of
consequential {\it nonunitarity} type as a necessary condition to exit
from the equivalence class of RQM, thus requiring new mathematical methods.

%%%%%%%%%%%%%%%%%%%%%%%%%%%%%%%%%%%%%%%%%%%%%%%%%%%%%%%%%%%%%%%%%%%%%%%%%%%%%

\section{Relativistic hadronic mechanics}

The insufficiencies of RQM for the
nuclear structure (as well as for the structure of hadrons and stars here
ignored) have been recognized  by various physicists, resulting in a  number
of research lines, such as the well known $q$-{\it deformations, quantum
groups,} and others. Unfortunately, even though {\it mathematically}
impeccable, these deformations are afflicted by a number of problematic
aspects of {\it physical} character [3c], besides implying an evident
departure from the axioms of the special relativity (SR).

In this note we use a covering of RQM known as {\it relativistic hadronic
mechanics} (RHM), originally submitted and studied by Santilli [4] (for
independent studies see monographs [5] and papers quoted therein), which
apparently resolves the above problematic aspects, and permits quantitative
and invariant studies of nuclear structures with extended, nonspherical and
deformable nucleons under nonlinear, nonlocal and nonunitary nuclear
forces [4h].

RHM is constructed via axiom-preserving maps of RQM called {\it isotopies}
[4a] here referred to {\sl maps  of any given linear, local-differential and
unitary theory into its most general possible nonlinear, nonlocal-integral
and nonunitary extensions which are nevertheless capable of reconstructing
linearity, locality and unitarity on certain generalized spaces called}
{\it isospaces}, {\sl over certain generalized fields called}
{\it isofields.}
The alterations characterized by RHM are called {\it mutations} [6] in order
to distinguish them from the ``deformations'' of the current literature.

The representation of physical systems via  RHM requires [1c,4h] the
knowledge of {\sl two} quantities: the Hamiltonian $H$ for the representation
of all potential interactions; and the isotopic generalization of the
conventional unit $I=\mbox{\rm diag}(1,1,1,1)$ of RQM, called {\it isounit},
$\hat I =\mbox{\rm Diag}({n_1}^2,{n_2}^2,{n_3}^2,{n_4}^2)\times\hat{\Gamma}
(x,\dot x,\psi,\partial\psi ,\ldots )=\hat I^{\dagger}>0$, where
${n_1}^2,{n_2}^2,{n_3}^2$ represent {\sl extended, nonspherical and
deformable shapes of the nucleons} under the volume preserving condition
${n_1}^2\times{n_2}^2\times{n_3}^2=1$, ${n_4}^2$ represents the {\it
density} {\sl of the medium in which motion occurs} (e.g., the square of the
local index of refraction), and $\hat{\Gamma}$ represent all nonlinear,
nonlocal and nonpotential interactions.

The {\it isotopic lifting} of the basic (multiplicative) unit $I\rightarrow
\hat I$ requires, for consistency, the corresponding lifting of the
conventional associative product $A\times B$ of RQM into the {\it
isoproduct} $A\hat{\times}B=A\times\hat T\times B$, $\hat I=\hat T^{-1}$,
under which $\hat I $ is the correct left and right unit of the new theory,
$\hat I\hat{\times}A\equiv A\hat{\times}\hat I\equiv A$ [4a]. The
{\it totality} of the products of RHM must therefore be isotopic for
consistency.

\def\tim{\times}
\def\dag{\dagger}
\def\arr{\rightarrow}
\def\h{\hat}

A realization of RHM specifically conceived for the nuclear structure
is presented in the adjoining paper [1c]. In essence, RQM can be reconstructed
with respect to the new isounit $\hat I$ with isoproduct $A\hat{\times} B$
via the use of {\it nonunitary transforms}. In fact, for $U\times U^{\dagger}
\neq I$, the isounit is precisely given by the transform $I\rightarrow\hat I
=U\tim I\tim U^{\dag}=\hat I^{\dag}\:,$ the isoproduct is precisely given by
the transform $A\tim B\arr U\tim A\tim B\tim U^{\dag}=\hat A\tim\hat T\tim
\hat B=\hat A\hat{\tim}\hat B$, $\h A =U\tim A\tim U^{\dag}$,
$\h B =U\tim B\tim U^{\dag}$,
$\h T =(U\tim U^{\dag})^{-1}=\h I^{-1}$, the fundamental relativistic
commutation rules are subjected to the map $[p_{\mu},x^{\nu}]=
{\delta_{\mu}}^{\nu}\tim I\arr U\tim[p_{\mu},x^{\nu}]\tim U^{\dag}=
[{\h p}_{\mu}\h ,{\h x}^{\nu}]={\h p}_{\mu}{\h{\tim}}{\h x}^{\nu}-
{\h x}^{\nu}{\h {\tim}}{\h p}_{\mu}=-i{\delta_{\mu}}^{\nu}\times{\h I}$, etc.

\def\hi{\h I}
\def\p31{$\h P (3.1)$}

The reader should be aware that the above procedure implies: new field
$\h F$ (=$\h R$ or $\h C$ of real or complex) {\it isonumbers}
$\h n=n\tim\h I$ [4c]; new {\it isodifferential calculus} with
$\h d x^{\mu}=
{{\h I}^{\mu}}_{\nu}\tim dx^{\nu}$, $\h{\partial}/\h{\partial}x=
{{\h T}_{\mu}}^{\nu}\partial/\partial x^{\nu}$ [4g]; new {\it isolinear
momentum operator}  $\h p_{\mu}\h{\tim}|\h{\psi}>=-i\h{\partial}_{\mu}|
\h{\psi}>=-i{{\h T}_{\mu}}^{\nu}\partial_{\nu}|\h{\psi}>$; new {\it
iso-Hilbert space} $\h{\cal H}$ with inner product $<\h{\phi}|\tim{\h T}
\tim |\h{\psi}>\tim\hi
\in\h C$ and  normalization $<\h{\psi}|\tim{\h T}\tim |\h{\psi}>=I$; new
{\it isoeigenvalue equations} $\h H\h{\tim}|\h{\psi}>=\h E\h{\tim}|\h{\psi}>
\equiv E\tim |\h{\psi}>$, $E\in F$, $\h E=E\tim\hi\in\h F$; new {\it
Lie-Santilli isotheory} [5] based on the isoproduct $[\h A\h ,\h B]=
\h A\h {\tim}\h B-\h B\h {\tim}\h A$ [3b,4a]; new  {\it iso-Minkowski space}
[4e] $\h M$ over $\h R$ with isounit $\hi =\h T^{-1}$, isometric
$\h N_{\mu\nu}=\h{\eta}_{\mu\nu}\times\hi={\h T_{\mu}}^{\alpha}
(x,\dot x,\h{\psi},
\partial\h{\psi},\ldots )\tim\eta_{\alpha\nu}\tim\hi$, $\eta=\mbox{\rm diag}\,
(1,1,1,-1)$, and isoseparation $(x-y)^{\h 2}=[(x-y)^{\mu}\h{\eta}_{\mu\nu}
(x-y)^{\nu}]\tim\hi\in\h R$; a new image $\h P (3.1)$ of the {\Po}
symmetry first introduced by Santilli [4d-4f] under the name of
{\it iso-Poincar{\'e} symmetry}; new {\it isofunctional analysis} [5c,5e];
etc. For the reconstruction in isospaces over isofields of all axiomatic
properties of RQM, such as hermiticity, real-valuedness of Hermitean
operators, linearity, locality, unitarity, etc., we are forced to refer
the reader to ref. [1c,4h].

We merely recall here that $\h P (3.1)$ is the image of $P(3.1)$ for the
new unit $\h I\:.$ As such, \p31 is the universal invariance of the
isoseparation $(x-y)^{\h 2}$, that is, of a geometry whose unit
$\hi$ directly represents extended, nonspherical and deformable particles
under unrestricted, generally nonunitary external forces. As such,
computer visualization of {\p31} show the desired removal of the
Keplerian center, as assured by the presence of contact/zero-range
interactions.

We should also recall that RHM is characterized by new degrees of freedom
of RQM axioms which are given for constant $\hi =n^{2}$ by the Hilbert
space law $<\phi |\times |\psi>\times I\equiv <\phi |\times n^{-2}\times
|\psi >\times n^{2}=<\phi |\times\hi\times\psi >\times\hi$ with Minkowskian
counterpart $x^{2}=(x^{\mu}\times\eta_{\mu\nu}\times x^{\nu})\times I\equiv
(x^{\mu}\times (n^{-2}\times\eta_{\mu\nu})\times x^{\nu})\times n^{2}=
(x^{\mu}\times\h {\eta}_{\mu\nu}\times x^{\nu})\times\hi\:.$ Note that
these new invariances have remained undetected during this century because
they required the prior discivery of {\sl new numbers with arbitrary units}
$\hi$ [4c].

Also, RHM provides an explicit and concrete ``operator'' realization of the
``hidden variables'' $\lambda =\h T (x,\h\psi,\ldots)$ via the isoeigenvalue
equation $\h{H}\h{\tim}|\h{\psi}>=\h{H}\times\lambda\times |\h{\psi}>=
\h{E}\h{\tim}|\h{\psi}>=(E\tim\lambda^{-1})\tim\lambda\tim |\h{\psi}>=
E\tim |\h{\psi}>\:.$ As such, RHM is a form of ``completion'' of RQM much
along the celebrated argument by Einstein, Podolsky and Rosen for which
von Neumann's theorem, Bell's inequalities and all that do not apply owing
to the underlying {\it nonunitary} structure (see [1c,4h] for details).

An important property is that (for positive-definite isounits) all isotopic
structures are locally isomorphic to the original ones,
$\hat{F}\approx F$,
$\hat{\cal H}\approx {\cal H}$,
$\hat{M}\approx M$,
$\hat{P}(3.1)\approx P(3.1)$,
etc., and they coincide at the abstract, realization-free level by conception
and construction. Possible criticisms on the axiomatic structure of RHM
(generally due to lack of sufficient technical knowledge of this new field)
are therefore criticisms on the axiomatic structure of RQM.
In fact, RHM {\sl is not} a new mechanics, but merely a {\sl new realization}
of the abstract axioms of RQM, although the two realizations are
{\sl physically inequivalent} because connected by nonunitary transforms.

The reader should be aware that RHM admits a hierarchy of realizations to
represent a hiererchy of physical conditions of increasing complexity,
ranging
from the minimal conditions of mutual overlapping of hadrons in the nuclear
structure, to their maximal conditions of mutual penetration in the interior
of collapsing stars. The realization needed for nuclear physics is
constructed for the first time in the adjoint paper [1c], and it is
conceived to {\sl preserve conventional quantum laws}. In fact, we have the
conventional uncertainties $\Delta\h{r}\Delta\h{p}\geq{1\over 2}<
[\h{p},\h{x}]={1\over 2}<\h{\psi}|\tim\h{T}\tim\h{I}\tim\h{T}|\h{\psi}>/
<\h{\psi}|\tim\h{T}\tim|\h{\psi}>={1\over 2}\:(\hbar =1);$ the conventional
spin
${1\over 2}$ and related Pauli's exclusion principle; and other conventional
laws. Note that this implies the preservation by RHM of causality under
nonlocal interactions (because they are embedded in the unit), the validity
of the superposition principle for a highly nonlinear theory (because of the
reconstruction of linearity in isospace), the invariance under nonunitary
time evolutions (because they are reduced to the isounitary law
$W=\h{W}\tim\h{T}^{1/2}\,,\: W\tim
W^{\dag}\equiv\h{W}\h{\tim}\h{W}^{\dag}=
\h{W}^{\dag}\h{\tim}\h{W}=\hi\neq I$), and other features [1c,4h] which are
impossible for the conventional formalism of RQM.

Above all, {\sl RHM is based on the requirement of preserving the axioms
of the SR and merely realize them in isospace $\h M$ under the isosymmetry}
$\h P$ (3.1) [4]. This permits to extend the applicability of the SR
from the current restriction to elm waves or perfectly spherical and rigid
bodies moving in vacuum with maximal speed $c_0$, to elm waves or
nonspherical-deformable bodies  moving within physical media with maximal
speed $c=c_0/n_4$. In particular, the maximal causal speed on $\h M$
remains  $c_0$ because in the fourth component of the isotopic line element
we have the lifting ${c_0}^2\arr c^2={c_0}^2/{n_4}^2$, with the {\it inverse}
lifting of the related unit $I_{44}=I\arr \hi_{44}={n_4}^2$, under which
the conventional value $c_0$ is invariant, thus rendering the SR universal
[4h].

Nowadays RHM possesses several preliminary, yet numerical and significant
verifications in nuclear physics, particle physics, astrophysics,
superconductivity, biology and other fields which we cannot possibly review
here for lack of space [4,5].

%%%%%%%%%%%%%%%%%%%%%%%%%%%%%%%%%%%%%%%%%%%%%%%%%%%%%%%%%%%%%%%%%%%%%%%%%%%

\section{Exact representation of total nuclear\hfil\break magnetic moments}

As indicated earlier, total nuclear magnetic moments still lack an exact
representation via RQM after three-quarter of a century of studies. For
instance, for the deuteron we have the value
${\h{\mu}_{D}}^{\mbox{\scriptsize\rm exp}}=0.857$,
while: quantum mechanics yields the value
$\mu_{\mbox{\scriptsize\rm QM}}^{\mbox{\scriptsize\rm theor}}=0.880$
which is 2.6 \% off {\sl in excess} [2a]; all possible corrections via
RQM still remain with about 1 \% off [2b]; and the problem does not appear
to be solvable via quark models because the quark  orbits are too small
to yield the needed deviation.

The most plausible explanation of the above occurrence was  formulated
by the Founding Fathers of nuclear physics in the late 1940's [2a]. Recall
that nucleons are not point like, but have extended charge distributions
with the radius of about  1 fm. Since perfectly rigid bodies do not exist
in nature, the above ``historical hypothesis'' (as hereon referred to)
assumes that such distributions can be deformed under sufficient external
forces. But the deformation of a charged and spinning sphere implies a
necessary alteration of its intrinsic magnetic moment. In turn, this
permits the exact representation of total nuclear magnetic moments as
shown below.

Our fundamental assumption is that the exact representation of total nuclear
magnetic moments requires the lifting RQM $\arr$ RHM, with basic lifting
$P(3.1)\arr\h P(3.1)$. As recalled earlier, the intrinsic characteristics
of nucleons are perennial and immutable under $P(3.1)$, while $\h P(3.1)$ has
been constructed precisely to represent their alterability under sufficient
conditions.

One of the first experimental verifications of RHM is then the exact-numerical
representation of total nuclear magnetic moments. It was presented for the
first time in ref. [6] under a joint mutation of angular momentum and spin.
In this note we give a new representation under mutated magnetic moments
but conventional values of angular momentum and spin.

\def\sp{\mbox{\scriptsize\rm spin}}
\def\orb{\mbox{\scriptsize\rm orb}}
\def\tot{\mbox{\scriptsize\rm tot}}
\def\dg{\mbox{\rm Diag}}
\def\thr{\mbox{\scriptsize\rm theor}}

The fundamental equation of RHM needed for quantitative treatment of the
historical hypothesis is the isotopic image of Dirac's equation, called
{\it iso-Dirac equation} [4h]. It is characterized by a nonunitary image of
the conventional equation and can be written for total 6-dimensional
isounits $\hi_{\tot}=\hi^{\orb}\times\hi^{\sp}$,
$\hi^{\orb}=\dg\, ({n_1}^2,{n_2}^2,{n_3}^2,{n_4}^2)\times\h{\Gamma}=
\dg\, (\hi_{11},\hi_{22},\hi_{33},\hi_{44})$, $\hi^{\sp}=U^{\sp}\times
U^{\dagger\sp}\neq\hi$ (see [1c] for details)

\def\beg{\begin{equation}}
\def\en{\end{equation}}
\def\begr{\begin{eqnarray}}
\def\enr{\end{eqnarray}}
\def\tim{\times}

$$
[\h N^{\mu\nu}\h{\times}^{\orb}\h{\gamma}_{\mu}\h{\times}^{\sp}
(\h p_{\nu}-\h i\h{\times}\h e\h{\times}\h A_{\nu})-i\times\h m^{\h 2}]
\h{\tim}^{\orb}\h{\psi}=0\:, \eqno{(1a)}
$$
$$
\{\h{\gamma}_{\mu}\h ,\h{\gamma}_{\alpha}\}=
{\h{\gamma}}_{\mu}\tim\h T^{\sp}\tim
\h{\gamma}_{\alpha}+\h{\gamma}_{\alpha}\tim\h T^{\sp}\tim\h{\gamma}_{\mu}=
2\h{\eta}_{\mu\alpha}\tim\hi^{\sp}\:,\eqno{(1b)}
$$
$$
\h{\gamma}_{\mu}=
({\h T_{\mu\mu}}^{\orb})^{1/2}\tim U^{\sp}\tim\gamma_{\mu}\tim
U^{\dagger\sp}\tim\hi^{\sp}\:,\eqno{(1c)}
$$

\def\no{\noindent}

\no where $\h i\h{\tim}\h e\h{\tim}\h A_{\mu}=(i\tim e\tim A_{\mu})
\tim\hi$ and the elm potential $A_{\mu}$ is conventional, being
external and long range.

As one can see, the above equation represents particles, which: a) are
extended and deformable with all infinitely possible ellipsoidical
shapes with semiaxes ${n_{1}}^{2}\,,\:{n_{2}}^{2}\,,\:{n_{3}}^{2}$ (under the
volume preserving condition ${n_{1}}^{2}\tim {n_{2}}^{2}\tim {n_{3}}^{2}=1)
\:;$
b) propagating within a physical medium with index of refraction $n_{4}$
(of value generally different than 1); and c) under conventional
electromagnetic interactions plus unrestricted external forces (represented
by $\hat{\Gamma}$). When a system is considered from the exterior, all
nonlocal-nonpotential internal effects must evidently be averaged into
constants (due to their short range character), as it is the case for total
nuclear magnetic moments. This yields a mere isorenormalization of the
$n_{\mu}$'s hereon tacitly implied.

It is easy to see that iso-Dirac equation (1) preserves the conventional
eigenvalues of angular momentum and spin  due to its very construction via
nonunitary transform of conventional equations (see [1c] for details),
while providing the desired {\it mutation of the magnetic moment of
nucleons} [4b,6,1b]

$$
\h{\mu}_N=\mu_N\tim n_4/n_3\:,\quad N=n\mbox{ or }p\:.\eqno{(2)}
$$

The application to the {\sl exact} representation of total nuclear magnetic
moments is straightforward. Assume to a good approximation that protons and
neutrons have the same shape $(n_{kn}=n_{kp})$ and that they  move in the same
medium $(n_{4n}=n_{4p})$. Then, a simple isotopy of the QM model [2a] yields
the {\it RHM model for the total nuclear magnetic moments}

$$
\hat{\mu}^{\tot}=\sum_{k} ({\h{g}_{k}}^{L}\tim\h{M}_{k3}+
{\h{g}_{k}}^{S}\tim\h{S}_{k3})\:,\eqno{(3a)}
$$
$$
\hat{g}_{n}=g_{n}n_{4}/n_{3}\approx g_{n}/n_{3}\:,\quad
\hat{g}_{p}=g_{p}n_{4}/n_{3}\approx g_{p}/n_{3}\:,\eqno{(3b)}
$$

\no where $e\hbar /2m_pc_0=1$, ${g_n}^S=-3.816$, ${g_p}^S=5.585$,
${g_n}^L=0$, ${g_p}^L=1$. As an illustration, the above model yields the
following {\it exact representation of the deuteron magnetic moment}

$$
\h{\mu}_{\thr}^{\tot}=
g_pn_{4p}/n_{3p}+g_n n_{4n}/n_{3n}\approx
(g_p+g_n)n_4/n_3\equiv{\mu_D}^{\exp}=0.857\:,\eqno{(4a)}
$$
$$
{n_4}^2=1.000\:,\quad {n_3}^2=1.054\:,\quad {n_1}^2={n_2}^2=(1/{n_3}^2)^{1/2}
=0.974\:.\eqno{(4b)}
$$

As one can see, ${{\mu}_D}^{\exp}$ is exactly represented by merely assuming
that the charge distribution of the nucleons in the deuteron experiences
a deformation of shape of about 1/2 \%. Note that the mutation is of
{\sl prolate} character which implies a {\sl decrease} of the (absolute value
of the) intrinsic magnetic moment of nucleons exactly as needed. Note also
that the representation is of {\sl geometric} character; it is independent
from any assumed nucleon constituent; and it identifies the polarization
of the constituent orbits which is needed for their compliance with physical
reality. Corrections due to the value $n_4\neq 1$ for the deuteron are
of 2-nd or higher order  (due to the relatively large  nucleon distace in the
deuteron).

The application of the model (3) to the exact representation of the total
magnetic moment of tritium, helium and other nuclei is straightforward and
studied in a future work.

Note that, the mutation of the charge distribution of the nucleons
{\sl is not} a universal constant, because it depends on the local conditions,
thus being generally different for different nuclei. This illustrates
the need of having infinitely possible {\sl different} isounits $\hi$.

%%%%%%%%%%%%%%%%%%%%%%%%%%%%%%%%%%%%%%%%%%%%%%%%%%%%%%%%%%%%%%%%%%%%%%%%%%%%

\section{The basic principle for possible new\hfil\break
recycling of nuclear waste}

RHM in its nuclear realization [1c] and the fundamental iso-Poincar{\'e}
symmetry
[4c-4f] predict the possible mutation not only of the intrinsic magnetic
moment of the neutron, but also of its meanlife, to such an extent that the
former implies the latter and {\it vice versa} (as one can see via the use of
the isoboosts). In turn, the control of the meanlife of the
neutron {\it de facto} implies new means for recycling the nuclear waste.

In this respect, the first physical reality which should be noted and
admitted is that total nuclear magnetic  moments constitute {\it experimental
evidence} on the alterability of the intrinsic magnetic moments of nucleons.

The second physical reality which should be noted and admitted is that,
by no means, the neutron has a constant and universal meanlife, because it
possesses a meanlife of the order of seconds when belonging to certain
nuclei with rapid beta decays, a meanlife of the order of 15 minutes when in
vacuum, a meanlife of the order of days, weeks and years when  belonging to
other nuclei, all the way to an infinite meanlife for stable nuclei.

Once the above occurrences are admitted, {\sl basic principle for possible
new recycling of nuclear waste is the} {\it ``stimulated neutron decay''}
(SND) {\sl consisting of resonating or other subnuclear mechanisms suitable
to stimulate its beta decay.} Among the various possibilities under study,
we quote here the possible {\it gamma stimulated neutron decay} (GSND)
according to the reaction [7b]

$$
\gamma+n\arr p^++e^-+\bar{\nu}\:,\eqno{(5)}
$$

\no which is predicted by RQM to have a very small (and therefore practically
insignificant) cross section as a function of the energy, but which is instead
predicted by RHM to have a resonating peak in said cross section at the
value of 1.294 MeV (corresponding to $3.129\times 10^{20}$ Hz). As such, the
above mechanism is of {\it subnuclear} character, in the sense of occurring
in the {\it structure  of the neutron}, rather than in the nuclear structure,
the latter merely implying possible refinements of the resonating frequency
due to the  (relatively smaller) nuclear binding energy [7b].

When stable elements are considered, the above GSND is admitted only in
certain instances, evidently when the transition is compatible  with all
conventional laws. This is the case for the isotope Mo(100,42) which, under
the GSND, would transform via beta emission into the Te(100,43) which, in
turn, is naturally unstable and beta decays into  Ru(100,44). For a number
of additional  admissible elements see [7b].

The point important for this note is that the GSND is predicted to be
admissible for large and unstable nuclei as occurring in the nuclear waste.
The possible new form of recycling submitted for study in this note is given
by {\sl  bombarding the radioactive waste with a beam of photons of the needed
excitation frequency and of the maximal possible intensity.} Such a beam
would cause an instantaneous excess of peripheral protons in the waste
nuclei with their consequential decay due to instantaneous excess of
Coulomb repulsive forces.

It should be stressed that this  note  can only address the basic {\sl
principle} of the GSND. Once experimentally established (see next section),
the recycling requires evident additional technological studies on the
{\sl equipment} suitable to produce the photon beam in the  desired
frequency and intensity, eg., via synchrotron radiation or other mechanisms
[7e].

The important point is that equipment of the above nature is expected to be
definitely smaller in size, weight and cost than large  particle accelerators.
As such, the recycling is expected to verify  the basic requirement of
usability by the nuclear power companies in their own plants.

A novelty of this note is that the study of recycling mechanisms is
specifically restricted to the {\it subnuclear} level. A virtually endless
number of possibilities exist for the reduction of the meanlife of the
waste via mechanisms of {\it nuclear} type. Among them we note mechanisms
based on RQM, such as those by  Shaffer et al. [8a], Marriot et al. [8b],
Barker [8c] and others, as well as new {\it nuclear} mechanisms predicted
by RHM and currently under additional patenting. The understanding is that,
to maximize the efficiency, the final equipment is expected to be a
combination of various means of both subnuclear and nuclear character.

\def\m{\mbox}
\def\res{\mbox{\scriptsize\rm res}}
\def\stim{\mbox{\scriptsize\rm stim}}
\def\spont{\mbox{\scriptsize\rm spont}}

As a final comment, the reader should be aware that any new recycling of
nuclear
waste is unavoidably linked to possible new sources of energy. In fact,
the GSND $\gamma_{\res}+\m{Mo}(100,42)\stackrel{\stim}{\arr}\m{Tc}(100,43)+
\beta\stackrel{\spont}{\arr}\m{Ru}(100,44)+\beta$ is {\it de facto}
a potential new source of {\it subnuclear} energy (releasing the rather
large amount of about 5 MeV plus the energy of the  $\gamma_{\res}$
per nucleus) called {\it hadronic energy} [7b-7g]. It should be noted that,
it confirmed, the new energy would not release harmful radiations, would
not imply radioactive waste, would not require heavy shield or critical
mass, and would be realizable in large or minuterized forms, thus having
ample prerequisites for additional studies.

%%%%%%%%%%%%%%%%%%%%%%%%%%%%%%%%%%%%%%%%%%%%%%%%%%%%%%%%%%%%%%%%%%%%%%%%%%%%%%

\section{Needed basic experiments}

The continuation of quantitative
scientific studies on the proposed new recycling of the nuclear waste
(as well as on possible subnuclear forms of clean energy) beyond the level
of personal views one way or another, requires the following three basic
experiments, all of moderate cost and fully realizable with current
technology.

\vspace{0.5\baselineskip}

{\bf 1. Finalize the interferometric $4\pi$ spinorial symmetry measures [9].}
Preliminary direct experimental measures on the alterability
of the {\it intrinsic} magnetic moments of nucleons were conducted from
1975 to 1979 by H. Rauch and his associates  [9a-9e] via interferometic
measures of the {\it 4$\pi$ spinorial symmetry} of the neutron. The best
available measures [9e] dating back to 1979 indicate about 1\% deviations
from $720^{\circ}$. But such deviation is {\sl smaller} than the error, and
the measures are therefore undefined. Similar unsettled measures were
conducted by Werner and his associates [9f] in the mid 1970's.

The above measures are manifestly fundamental for possible new forms of
recycling as well as for possible new forms of subnuclear energy. In fact,
they would provide experimental evidence on possible deviations from  the
Poincar{\'e} symmetry in favor of our covering iso-Poincar{\'e} form [4].
This is due to the fact that, if confirmed, the measures would establish a
deviation from the fundamental {\it spinorial} transformation law of favor
of the mutated form easily derivable from Eqs. (1) (first identified in [6],
see [1c] for a recent derivation)

$$
\h{\psi}'=\h R(\theta_3)\h{\tim}\h{\psi}=e^{i\gamma_1\gamma_2\h{\theta}_3/2}
\tim\h{\psi}\:,\quad\h{\theta}=\theta/n_1\tim n_2\:,
\eqno{(6)}
$$

As an illustration, assume a 1\% deviation from $720^{\circ}$. The isotopies
re-construct the exact SU(2)-spin in isospace, thus requiring
$\h{\theta}=\theta/n_1\tim n_2 =720^{\circ}$ [4h]. This yields

$$
{n_1}^2={n_2}^2=713^{\circ}/720^{\circ}=0.990\:,\quad{n_3}^2=1.020\:,
\eqno{(7a)}
$$

$$
\h{\mu}/\mu=n_4/n_3\approx 713^{\circ}/720^{\circ}\:,\quad
n_4=n_3\tim 713^{\circ}/714^{\circ}=1.000\:,\eqno{(7b)}
$$

\no namely, our iso-Dirac equation provides an {\sl exact-numerical}
representation of the $4\pi$ interferometric measures, by {\sl deriving}
the value $n_4=1$ occurring in vacuum, as it is the case for the neutrons
of tests [9] (but only approximately so for the deuteron).  For more details
see [1c].

Once, in addition to the evidence from total nuclear magnetic moments
and various other applications [4,5], the validity of the iso-Poincar{\'e}
symmetry is established via direct experiments for the mutation of the
intrinsic magnetic moment of the neutron, that of its meanlife is expected
to be consequential.

\vspace{0.5\baselineskip}

{\bf 2. Repeat don Borghi's experiment [7f] on the apparent
synthesis of the neutron from protons and electrons {\sl only}.}
Despite momentous advances, we still miss fundamental experimental knowledge
on the structure of the neutron, e.g., on how the neutron is synthesized
from protons and electrons {\sl only} in young stars solely composed
of hydrogen (where quark models cannot be used owing to the lack of the
remaining members of the baryonic octet, and  weak interactions do not
provide sufficient  information on the {\it structure} problem).

The synthesis occurs  according to the reaction $p^++e^-\arr n+\nu$
which: is the ``inverse'' of  the stimulated decay (5); is predicted
by RQM to have a very small cross section as a function of the energy;
while the same cross section is predicted by RHM to have a peak at the
threshold energy of 0.80 MeV in singlet $p$-$e$ coupling [4f].

The possible synthesis of the neutron has a fundamental relevance for waste
recycling, besides other industrial applications. If the electron
``disappears'' at the creation of the neutron as in current theoretical
views,  the GSND becomes  of  difficult if not impossible realization.

However, the electron is a permanent and stable particle. As such, doubts
as to whether it can ``disappear'' date back to Rutherford's very conception
of the neutron as a ``compressed hydrogen atom'' [7a]. As well known, RQM
does not permit such a representation of the neutron  structure  on numerous
counts. Nevertheless, the covering RHM has indeed  achieved an exact-numerical
representation of all characteristics of the neutron according to
Rutherford's original conception [4f]. The novelty is that, when immersed
within the hyperdense proton, the electron experiences  a  mutation
$e^-\arr\h e^-$ of its {\it intrinsic} characteristics (becoming a quark ?)
including its rest energy (because $n_{4}\neq 1$ inside the proton, thus
$E_{\h e}=mc^{2}+m_{e}{c_{0}}^{2}/{n_{4}}^{2})\:.$ The excitation energy of
1.294 MeV is predicted by our covering iso-Poincar{\'e} symmetry under the
condition of recovering all characteristics of the neutron for the model
$n=({p^{+}}_{\uparrow},{{\h e}^{-}}_{\downarrow})_{\mbox{\scriptsize\rm RHM}}
\:,$ including its primary decay for which $\h{e}^{-}\rightarrow e^{-}+
\bar{\nu}$ [4f].

A preliminary experimental verification of the synthesize the neutron in
laboratory was done by don Borghi's and his associates [7d]. Since
experiments can be confirmed or dismissed {\sl solely} via other experiments
and certainly not via theoretical beliefs one way or the other, don Borghi's
experiment must be run again. The test can be repeated either as originally
done [7f], or in a number of alternative ways, e.g., by hitting with a
cathodic ray of 0.80 MeV a mass of beryllium saturated with hydrogen, put
at low temperature and subjected to an intense electric field to maximize
the $p$-$e$ singlet coupling. The detection of neutrons emanating from such
a set-up would establish their synthesis.

\vspace{0.5\baselineskip}

{\bf 3. Complete Tsagas' experiment [7g] on the stimulated neutron decay}.
The last and most fundamental information needed for additional
quantitative studies is the verification or disproof of the GSND at the
resonating gamma frequency of 1.294 MeV.

The latter experiment has been initiated by N. Tsagas and his associates
[7g]. It consists of a disk of the radioisotope Eu$^{152}$ (emitting gammas
of 1.3 MeV) placed parallel and close to a disk of an element admitting
of the GSND, such as the Mo(100,42) (or a sample of nuclear waste).
The detection of electrons with at least 2 MeV emanating from the system
would establish the {\sl principle} of the GSND (because such electrons can
only be of subnuclear origin, Compton electrons being of at most 1 MeV).
The detection via mass spectrography of traces of the extremely rare
Ru(100,44) after sufficient running time would confirm said principle.
The practical realization of the proposed form of waste recycling would then
be shifted to the industrial development and production of a photon beam of
the needed frequency and intensity.

%%%%%%%%%%%%%%%%%%%%%%%%%%%%%%%%%%%%%%%%%%%%%%%%%%%%%%%%%%%%%%%%%%%%

%%%%%%%%%%%%%%%%%%%%%%%%%%%%%%%%%%%%%%%

\end{document}